\PassOptionsToPackage{dvipsnames}{xcolor}
\documentclass[sigconf]{acmart}

\AtBeginDocument{%
  \providecommand\BibTeX{{%
    \normalfont B\kern-0.5em{\scshape i\kern-0.25em b}\kern-0.8em\TeX}}}

\setcopyright{none}

\acmConference[CIDR'22]{}{Chaminade CA}{USA}

\usepackage{mathptmx}
\usepackage{graphicx}
\usepackage{times}
\usepackage{listings}
\usepackage{courier}
\usepackage{svg}
\usepackage{subcaption}
\usepackage{balance}
\usepackage{amsmath}
\usepackage{url}
\usepackage{xspace}
\usepackage{pifont}
\usepackage{outlines}
\usepackage{multirow}
\usepackage{float}
\usepackage{comment}

\newif\ifnotes
\notestrue

\newcommand{\subhead}[1]{\vspace{1ex}\noindent\textbf{#1}}

\newcommand{\system}{{\sc SigmaTyper}\xspace}
\newcommand{\type}[1]{\texttt{#1}}

\renewcommand\footnotetextcopyrightpermission[1]{}

\begin{document}

\title{Making Table Understanding Work in Practice}

\author{Madelon Hulsebos}
\affiliation{%
  \institution{Sigma Computing}
  \city{San Francisco}
  \country{USA}
}
\email{madelon@sigmacomputing.com}
\additionalaffiliation{%
  \institution{University of Amsterdam}
  \city{Amsterdam}
  \country{Netherlands}
}

\author{Sneha Gathani}
\affiliation{%
  \institution{Sigma Computing} 
  \city{San Francisco}
  \country{USA}
}
\email{sneha@sigmacomputing.com}
\additionalaffiliation{%
  \institution{University of Maryland}
  \city{College Park}
  \country{USA}
}

\author{James Gale}
\affiliation{%
  \institution{Sigma Computing}
  \city{San Francisco}
  \country{USA}
}
\email{jlg@sigmacomputing.com}

\author{Isil Dillig} 
\affiliation{%
  \institution{University of Texas}
  \city{Austin}
  \country{USA}
}
\email{isil@cs.utexas.edu}

\author{Paul Groth}
\affiliation{
  \institution{University of Amsterdam}
  \city{Amsterdam}
  \country{Netherlands}
}
\email{p.t.groth@uva.nl}

\author{\c{C}a\u{g}atay Demiralp}
\affiliation{%
  \institution{Sigma Computing}
  \city{San Francisco}
  \country{USA}
}
\email{cagatay@sigmacomputing.com}

\begin{abstract}
Understanding the semantics of tables at scale is crucial for tasks like data integration, preparation, and search. Table understanding methods aim at detecting a table's topic, semantic column types, column relations, or entities. With the rise of deep learning, powerful models have been developed for these tasks with excellent accuracy on benchmarks. However, we observe that there exists a gap between the performance of these models on these benchmarks and their applicability in practice. In this paper, we address the question: what do we need for these models to work in practice?

We discuss three challenges of deploying table understanding models and propose a framework to address them. These challenges include 1) difficulty in customizing models to specific domains, 2) lack of training data for typical database tables often found in enterprises, and 3) lack of confidence in the inferences made by models. We present \system which implements this framework for the semantic column type detection task. \system encapsulates a hybrid model trained on GitTables and integrates a lightweight human-in-the-loop approach to customize the model. Lastly, we highlight avenues for future research that further close the gap towards making table understanding effective in practice.
\end{abstract}

\maketitle

\begingroup

\section{Introduction}
The data used and generated by enterprise applications today are processed and stored overwhelmingly as relational tables in relational databases, including data warehouses. Table understanding aims at surfacing the semantics of tables in order to improve and help automate many tasks from data integration to data visualization. Recent developments in machine learning (ML), particularly deep representation and transfer learning, suggest that the time might be ripe for augmenting enterprise applications with computational table understanding. Table understanding~\cite{tableunderstandingtutorial2021} in this context refers to tasks such as topic inference, table annotation, semantic column type detection, entity resolution, relation extraction, and learning contextual semantic representations of table elements.

Surfacing table semantics as metadata and effective representations (e.g., contextual semantic embeddings) can improve data exploration tasks such as search and data discovery in applications~\cite{webtablestenyears}. 
For example, knowledge about semantic column types and inter-column relations is  useful for automating  data preparation~\cite{chu2015katara} and data visualization~\cite{hu2019vizml,dibia2019data2vis,zhou2020table2charts}. Knowledge of table schemas and entities, in turn, facilitates the integration of two or more data sources~\cite{stonebraker2018dataintegration} and can be used to construct data catalogs.

Recent work has introduced deep learning models such as Sherlock~\cite{hulsebos2019sherlock} and SATO~\cite{zhang2020sato} to detect semantic column types. Using transformers~\cite{vaswani2017attention} and building on BERT~\cite{devlin2019bert}, researchers have also proposed pretrained table models such as TURL~\cite{turl}, TaBERT~\cite{yin2020tabert}, TaPaS~\cite{herzig2020tapas}, TABBIE~\cite{iida2021tabbie} and TUTA~\cite{wang2021tuta}
and successfully applied them in various downstream tasks for table understanding. While there is a heightened research interest in table understanding models, little is known about the feasibility of these 
models in practice.

Commercial data systems such as Talend~\cite{talend}, Trifacta~\cite{trifacta}, and Tableau~\cite{tableau}, in general, do not seem to deploy such models in practice. 
Instead, they primarily rely on simpler methods like regular expression matching for detecting a limited set of semantic types~\cite{trifacta:type-system, talend:type-system,  googledatastudio:semantic-type}. Tamr~\cite{stonebraker2013tamr} made the step towards machine learning but uses 
less data-heavy machine learning models due to the unavailability of training data and lack of domain expert time, among other things~\cite{stonebraker2018dataintegration}.

Motivated by the abundance of practical applications of table understanding, we shift our focus to making these models effective in practice. We identify three challenges in deploying table understanding models in practice.
First, early feedback from companies on semantic column type detection models such as Sherlock and SATO, highlighted the desire to customize models for different sets of tables and semantic types specific to domains. Doing so requires new large-scale datasets and semantic types that represent these domains well which often requires efficient labeling procedures and long retraining procedures. Although the pretraining/finetuning paradigm adopted for table understanding models~\cite{turl,yin2020tabert} is meant to relieve these burdens, we observe that these models are not straightforward to finetune and still require a significant set of labeled data.

Second, these models are pretrained on tables that poorly resemble typical database tables, as the training data mostly reflects tables found on the Web~\cite{langenecker2021towards,hulsebos2021gittables}. Therefore, tuning a pretrained model or retraining a task-specific model towards a representative data distribution and labels still takes many resources.
Lastly, the quality and confidence of the inferences made with these models are typically unstable across labels~\cite{hulsebos2019sherlock}, which limits the reliability of the model output in practice. Besides reliability, no finetuning procedures are proposed to reflect whether samples are far from the training set~\cite{dhamija2018reducing}. In practice, it is important to accurately reflect out-of-distribution data points to ensure high precision.

Based on these challenges, we propose a framework that provides a foundation for table understanding systems in practice. This framework is principled in enabling lightweight and interactive customization towards the target data domain, generalizing to typical database tables, and ensuring high precision and coverage. We present \system, a system that implements the proposed framework for the semantic column type detection task but generalizes to other table understanding tasks. 
\system incorporates a pretrained global model identically deployed across all customers, which combines heuristics with a learned model to establish high precision and semantic type coverage. This base model adapts iteratively to the user’s context based on user feedback. Building on the data programming by demonstration (DPBD) paradigm~\cite{evensen2020dataprogramming}, \system generates custom training data for the model leveraging the user's feedback. \system is pretrained on GitTables providing representative tables as well as relevant semantic types.

In summary, our contributions are as follows: 1) the identification of three key challenges of table understanding systems in practice; 2) a framework for building table understanding systems; and 3) \system, an implementation of this framework for semantic column type detection.

\section{Challenges}\label{sec:challenges-practice}
Deploying table understanding models introduces many challenges in practical applications. Here, we discuss three challenges deduced from our experience and the literature. 

\subsection{Unknown data domains} 
\begin{figure}
\hspace{-0.1cm}
\includegraphics[scale=0.48]{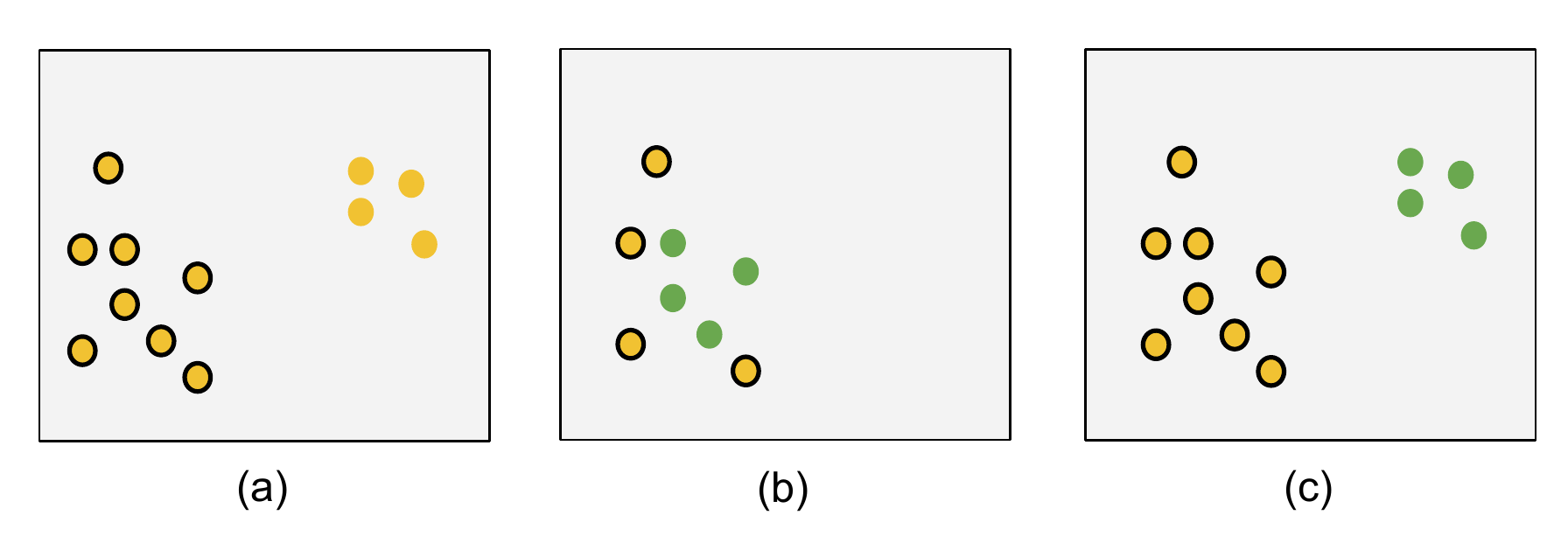}
\caption{Models often fail when data-shift occurs. Data-shift comes in three flavors in practice: (a) covariate-shift, (b) label-shift, and (c) out-of-distribution data. Circles represent sample data points and their colors indicate the labels of the samples. The border of a circle indicates whether the corresponding sample stems from the training set (black) or target set (none).}
\label{fig:adaptation-scenarios}
\end{figure}

One system does not fit every context. In order to provide responsive and intelligent functionalities, the accuracy of a table understanding system is expected to converge to a user's context and improve over time. If systems fail to do so, they might lose relevance to a user, leading to potential decreased satisfaction or engagement.

There are three cases of shift that can occur between the training data and the data in the user context, the inference-time data. The first case is when the distribution of table values present in the training data differs from that of the inference-time 
tables. This is also called covariate-shift~\cite{kouw2018transferlearning} (Fig.~ \ref{fig:adaptation-scenarios}a). For example, in the case of semantic column 
type detection, a column with the name ``ID'' might contain values not previously seen for the semantic type \type{ID}.
The second case is when table values associated with a certain label in the training dataset correspond to a different label within the user's context. This is also known as label-shift (Fig.~\ref{fig:adaptation-scenarios}b)~\cite{rabanser2018failing}.
For example, a column with predicted semantic type \type{ID} might actually correspond to the \type{phone number} type within the user's context. The third is the case when we encounter out-of-distribution data (Fig.~\ref{fig:adaptation-scenarios}c). A user might have tables and labels far from the training distribution~\cite{rabanser2018failing}.

\subsection{Unrepresentative training data} 
With the availability of large-scale table resources like WebTables~\cite{webtables2012} the fruits of deep learning were leveraged to build deep learning models for semantic column type detection models~\cite{hulsebos2019sherlock, zhang2020sato}. Succeeding methods extend the pretraining paradigm to develop task-agnostic pretrained table understanding models~\cite{yin2020tabert, turl, wang2021tcn}.

Despite the abundance of pretrained table understanding models that have been published, they seem not widely adopted in practice. Business Intelligence systems like Tableau and Google Data Studio, for example, only detect a limited set of semantic types, seemingly with lookup functions or pattern matching~\cite{trifacta:type-system, googledatastudio:semantic-type}.

We believe that this gap can be attributed to the datasets used to pretrain these models, which mainly represent tables from the Web. Such tables can only partially represent tables found in enterprise databases~\cite{hulsebos2021gittables,langenecker2021towards, vogelsgesang2018getreal}. This affects the applicability of concurrent pretrained table models to downstream tasks on typical ``offline'' databases. To remove this barrier, we need models pretrained on relevant tables, resembling what one encounters in offline databases.

\subsection{Unreliable inferences} 
A table understanding system is useless to users, and might even do more harm than good if it cannot accurately infer table semantics. For example, a column type detector should be able to accurately predict basic types such as \type{date} and \type{city} with high precision as a bare minimum, while more and specific semantics may be a nice-to-have. Balancing this precision with coverage, e.g. the number of labels a model is able to distinguish, is a hard challenge~\cite{deng2010imagecategories} but finding the optimal operating point is critical to make a table understanding system effective.

Finally, in practice, and especially in applications with varying semantic contexts, we might encounter tables and target labels that are not covered in the training set (Fig. \ref{fig:adaptation-scenarios}c). While high-capacity models exist, they do not extrapolate well and can hardly capture the entire universe. Upon encountering tables and labels that are far from the training data, the system should avoid inferring labels for it. This is mainly critical in exact or user-facing applications such as data integration, visualization recommendation, or data cataloging. 
If a system fails in this regard, we risk losing a user's trust in the system or produce critical errors. The question is, (how) do table understanding models recognize out-of-distribution tables? And how to handle situations where a model fails?
\begin{figure*}
    \centering
    \includegraphics[scale=0.43]{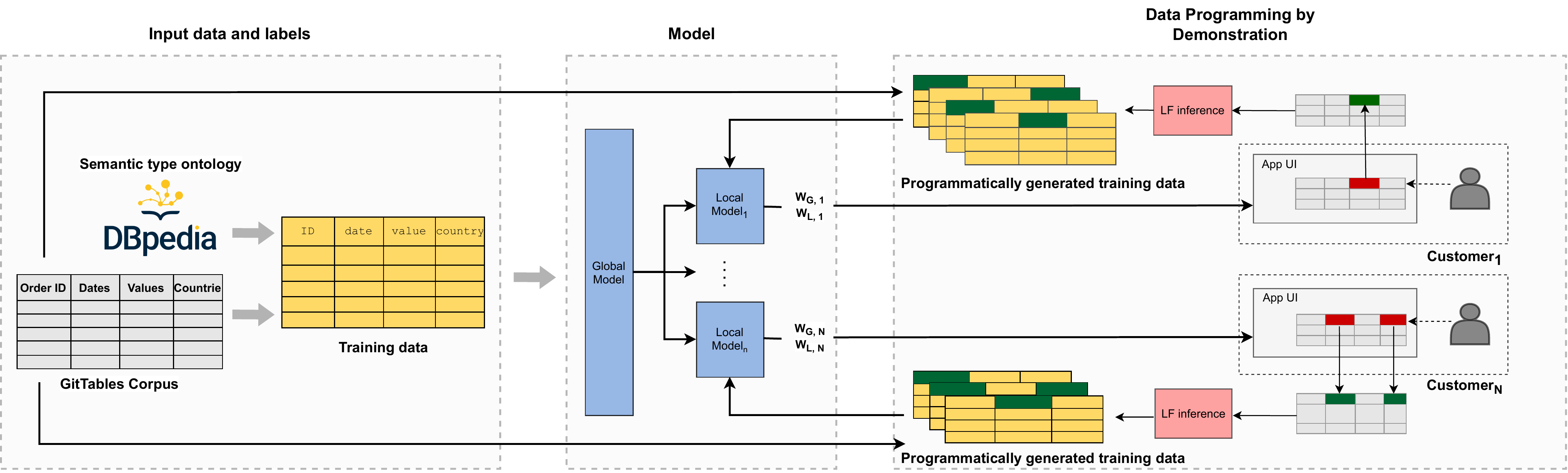}
    \caption{\system: tables and semantic types are used to pretrain a global model. The local model adapts towards the user's context through data programming by demonstration: based on the user's feedback, labeling functions are inferred and used to generate new training data and function as weak predictors in the local model. The weight of the local model increases over time.}
    \label{fig:system-overview}
\end{figure*}

\section{Framework}
We propose a general framework to addresses these challenges, providing a foundation for implementing table understanding 
systems in practice.

\subsection{Interactive customization}
Table understanding systems should effectively adapt to the user's context. Earlier information retrieval systems have shown that iteratively learning from interactive feedback is an efficient way to accomplish this~\cite{tamine2010evaluation, joachims2007search}.  Interactions for this end should be easy to use, taking minimal time and input, to maximize the effectiveness of the feedback cycle. Existing systems often, however, require complicated configurations~\cite{talend:type-system, googledatastudio:semantic-type}.

In our vision, the user's feedback, implicit or explicit, should be used to derive rules automatically (e.g. regular expressions or heuristics) following the programming by demonstration (or example) framework~\cite{lieberman2001your}. For example, if we want to match tables describing similar products from different e-commerce platforms, we can extract rules that describe the distribution of prices across products. Such rules can be exploited in many ways to adapt the system to the user's task and data context. We believe they can, at least, be used to 1) generate customized training data using data programming~\cite{ratner2016dataprogramming}, and 2) as weak learners to make new inferences~\cite{zhou2021ensemble}.

\subsection{Relevant training data}
Models are only as good as the data they are trained on. Unlike large corpora of text extracted from the Web which are shown to be instrumental for pretraining widely used language models~\cite{devlin2019bert, brown2020gpt3}, pretrained table models have shown less impact in this regard. In fact, the generalizability of models trained towards typical database tables is found to be limited~\cite{langenecker2021towards, hulsebos2021gittables}.

To deploy table understanding models in enterprises, we need training data relevant to a variety of domains. Typical corpora used for pretraining table models include WebTables~\cite{webtables2012} and WikiTables~\cite{bhagavatula2013wikitables}. Tables from the Web are relatively small and homogeneous. Typical database tables, instead, are relatively large and heterogeneous~\cite{vogelsgesang2018getreal}. Table understanding models should reflect such data distributions and be pretrained on matching corpora and relevant labels.

\subsection{Hybrid adaptive systems}
Systems should ensure high precision and high coverage, be assessed accordingly, and ensure reliability. We believe that systems should be hybrid, combining heuristics that are often pragmatic, fast, and transparent (e.g., easier to explain), with learned models that offer high capacity, flexibility, and semantic coverage.

Beyond accuracy and coverage, systems should learn to reflect their confidence about out-of-distribution samples by being trained specifically on these scenarios. Finally, the underlying pretrained models should adapt to new domains, and do so with minimal iterations.

\section{SigmaTyper}\label{sec:sigmatyper} 
We present \system, a system that implements the proposed framework for semantic column type detection 
(annotation), which is a popular table understanding task. Fig.~\ref{fig:system-overview} shows the high-level system architecture of our system. 

\subsection{Input data and labels}
As traditional large-scale table corpora do not extend well to database tables~\cite{langenecker2021towards, hulsebos2021gittables}, newer data sources~\cite{herzig2021open, hulsebos2021gittables, vogelsgesang2018getreal} aim at capturing more database-like tables. Since \system is intended to operate on enterprise tables, we use the GitTables~\cite{hulsebos2021gittables} corpus to train it.  GitTables has been recently introduced to address the need for database tables to train models for enterprise 
applications. 

Predictions of a learned model should be relevant for the majority of its intended users and their application domains. For semantic column type detection, this implies that we need semantic types common in the enterprise, science, and medical domains, and beyond. Tables in GitTables come along with over 500 semantic types from DBpedia~\cite{dbpedia} and Schema.org~\cite{guha2016schema}. We select the semantic types from the DBpedia ontology, given its broad semantic coverage and easy integration with the DBpedia Knowledge Base.

\begin{figure*}
    \centering
    \includegraphics[scale=0.48]{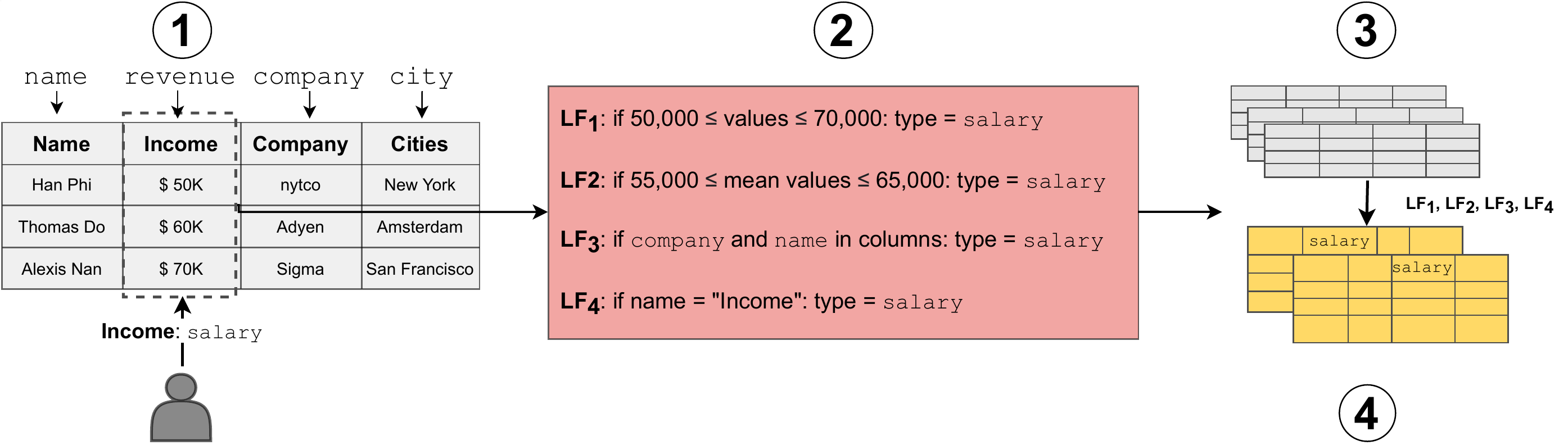}
    \caption{Data pogramming by demonstration implementation of \system. \ding{172} Based on the user's correction of ``Income'' to type \type{salary}, \ding{173} \system infers labeling functions (LFs) from the column using e.g. data profilers, and \ding{174} uses the LFs to extract customized training data from the source corpus, \ding{175} into customized weakly labeled training data for semantic type \type{salary}.}
    \label{fig:dpbd}
\end{figure*}

\subsection{System adaptation by DPBD}\label{sec:dpbd}
To stay useful, systems incorporating learned models should adapt to the user's specific semantic context, without imposing time-consuming work on the user. Hand-labeling vast amounts of data to retrain machine learning models, especially deep learning models, is infeasible given budget constraints~\cite{stonebraker2018dataintegration}. We build on the data programming by demonstration framework (DPBD)~\cite{evensen2020dataprogramming} to handle these challenges. That is, we infer labeling functions automatically from a table based on the user's feedback, which may be implicit or explicit, and use these functions to generate new training data.

Considering the example in Figure \ref{fig:dpbd}, the system yields initial column type predictions for each column in this table. The second column, labeled with type \type{revenue} is found to be incorrect by the user. The user explicitly relabels this column, originally named ``Income'', with new type \type{salary}. Labeling functions are then inferred for the new type \type{salary} and used to generate training data so that the model can be tuned to recognize the type in the future~\cite{ratner2016dataprogramming}. While the interface allows explicit approval of the type predictions, the user does not give this explicit feedback for the remaining columns but leaves those type predictions as-is and conducts an analysis on the table. The system automatically interprets the other types to be implicitly approved. The entire table with its labels is then added to the training data along with the newly generated training data.

\system infers labeling functions from columns with textual and numeric data types. For numeric columns, it captures statistics of the data distribution using a data profiler, currently Great Expectations~\cite{2020greatexpectations}, and extracts textual features from textual columns, e.g. the most frequent values and the number of unique values. For both type of columns, \system infers functions to indicate co-occurring columns based on the other detected types, e.g. the types of neighboring columns as in Figure~\ref{fig:dpbd}.

To accommodate customization without occluding the model for other customers, \system consists of a ``global'' and multiple ``local'' models, one for each customer. The newly generated training data is only used to adapt the local model. The influence of the global and local models on the final prediction is captured in weight vectors representing the influence of each model per type, i.e. $\textbf{W}_{g}$ for the global model and $\textbf{W}_{l}$ for the local model. Over time, the influence of the local model on the final prediction increases. 

\begin{figure}[tbh]
    \centering
    \includegraphics[scale=0.425]{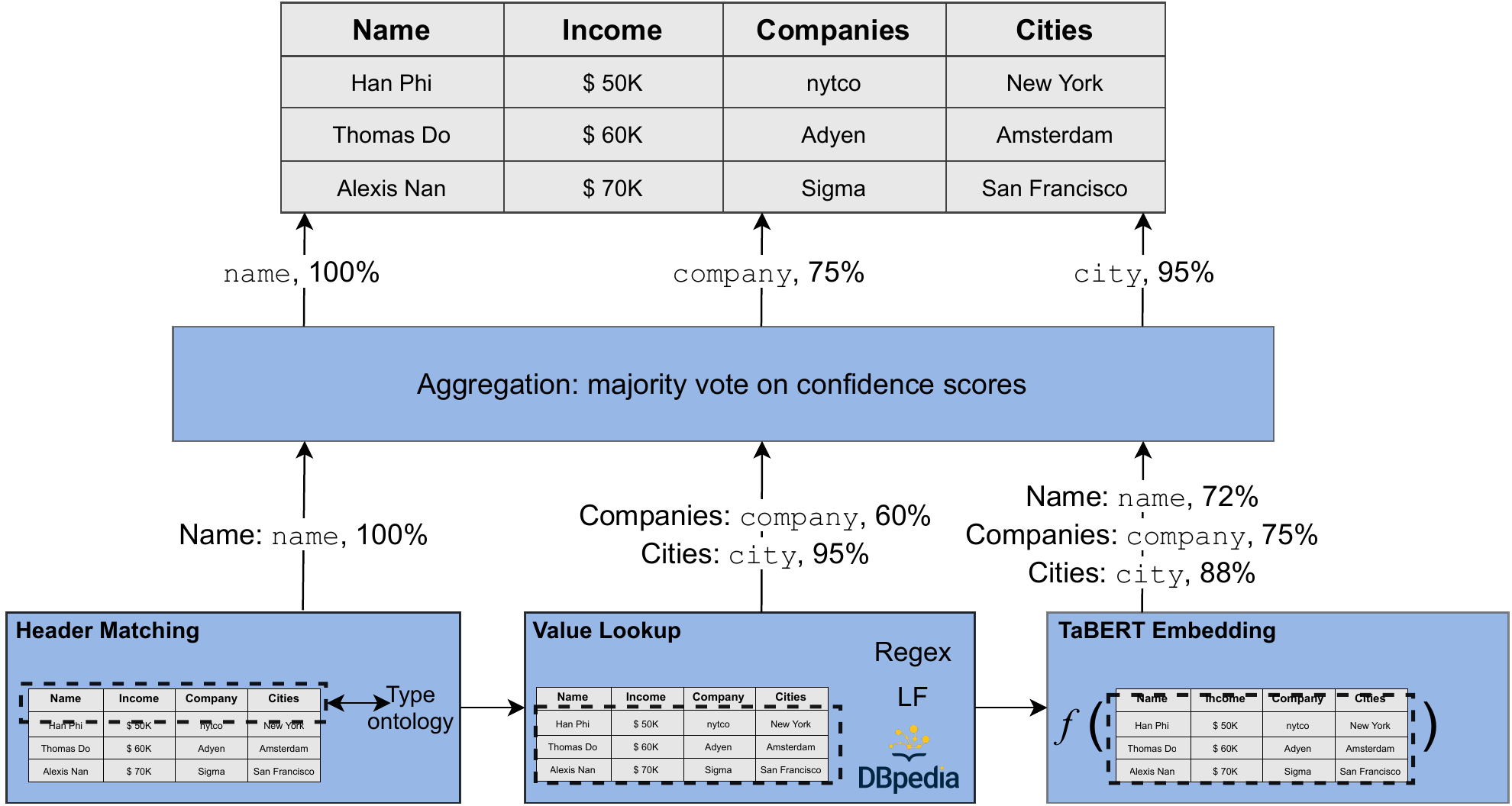}
    \caption{Semantic column type detection model of \system. It implements a 3-step pipeline to predict the semantic type of a column based on the column's header and values along with an embedding of the entire table.}
    \label{fig:detection-pipeline}
\end{figure}

\subsection{Semantic type detection pipeline}

\system implements a 3-step pipeline to predict the semantic types of columns based on the table header, column values, and an embedding of the entire table (Fig. \ref{fig:detection-pipeline}). 
To minimize overhead, each step in the pipeline is executed (potentially for a subset of columns) only if a preset confidence threshold $c$ is not met by the prior step. The steps are executed in order of inference time, so that the slowest step is executed at last. Given a table $T$ the system yields the top-$k$ semantic types for each column along with their confidence score.

As the header of $T$ might already correspond to a semantic type, \system matches each column name to the labels in the type ontology using syntactic and semantic matching. For syntactic matching, the column header is compared to the semantic types in the ontology using fuzzy matching. If there is a syntactic match, the confidence score is set to the maximum being $100\%$.For semantic matching, embedding vectors for the column names as well as the target semantic types are computed using FastText~\cite{joulin2016fasttext} first. Then the cosine similarity between the column names and semantic types are calculated using their respective embedding vectors. \system considers these similarity scores as prediction confidences and returns the target types with the highest confidences are returned as final predictions for each column.

If the confidence of the header matching step does not surpass a preset threshold for a subset of columns, a lookup step is triggered to label them. This step leverages lookup rules, to match a sample of column values to semantic types from the ontology. These rules consist of 1) the labeling functions from the global model and the local model which are obtained through DBPD (Sec. \ref{sec:dpbd}), 2) the DBpedia Knowledge Base~\cite{dbpedia}, and 3) a set of regular expressions which might be expanded on user input as well~\cite{chen2020multi}. The fraction of values that matched a type, is returned as the confidence for that type. The types with the highest confidences are returned as final predictions for each column.

If the confidence of prior steps for any column is low, the pipeline embeds $T$ using a pretrained TaBERT model~\cite{yin2020tabert}. We trained TaBERT's parameters towards GitTables and finetuned it to enable semantic column type detection. To detect out-of-distribution samples (Fig. \ref{fig:adaptation-scenarios}c) we train the model on a background dataset~\cite{reddy2019figurenet} and add the semantic type \type{unknown} to match columns of unknown types. The table embedding model in the local model is iteratively finetuned based on the weakly labeled data generated using DPDB. 

The final prediction for each column in $T$ is the soft majority vote based on the concatenated confidence scores from each step. An optimal aggregation function can be learned as well. We infer a parameter $\tau$ and threshold predictions that are below $\tau$ such that the precision of the system is high, hence errors are minimized.

\section{Future work}\label{sec:future-work}

\subhead{Evaluation.}
While a first prototype of the system is built, we aim at integrating it into a product over the next weeks. We plan to investigate if and how accurately the system addresses the challenges discussed in section~\ref{sec:challenges-practice}. In particular, we are interested in understanding how well and rapidly the system adapts to new data domains, how it generalizes to database tables and semantics in general, and in what scenarios it fails to make correct predictions.

\subhead{Adaptation from scratch.}
It may be necessary to tailor a model to a different table distribution or ontology from scratch. One possible route for adapting to new data distributions is to take inspiration from multilingual models and pretrain a model on samples from both ``languages''~\cite{lample2019crosslingual}, tables with different data distributions in our case. This model can be fine-tuned towards the downstream task on a set of labeled tables from the source training data, which in turn can be used to infer the labels for the target table distribution. For a different ontology, we could leverage alternative inexpensive data sources, from a different modality, labeled with the desired ontology~\cite{suri2020leveraging}.

\subhead{Engaging users to provide feedback.}
An ongoing research topic is how users can be engaged with systems to improve the system for collective reasons. Users need to have individual incentives in order to give explicit feedback, or it should be made very easy to give feedback. Preferably, a system does not even need any explicit feedback and learns from implicit feedback directly. We foresee many possibilities in this space and plan to experiment with users to learn what works best.

\subhead{Deeper analysis of table models.}
We also propose a further analysis of the boundaries of what these models can and cannot do to elicit further developments. Prior work reports model performances often measured by predictive accuracy and occasionally analyze on which labels they fail. But we want to structurally answer the question \textit{when} these models do not work, and \textit{why}. Analyzing these aspects, in-depth, will help understanding where the weaknesses of these models reside.

\subhead{Table-specific representations.}
To achieve further performance in practice, it seems necessary to move beyond adapting natural language models towards tables. TCN~\cite{wang2021tcn} and RPT~\cite{tang2021rpt} propose the first steps towards leveraging the structure of tables. TCN deploys a convolutional neural network, while RPT decouples a table into tuples of attribute rows. TCN confirms the value of developing methods to represent tables, as it demonstrates the largest performance lift compared to other methods that build on top of language models like BERT~\cite{devlin2019bert}. The latest model contribution, RPT, proposes a representation of row-wise tuples, solely leveraging row-wise relations. We believe these methods are a step in the right direction but in-depth analyses can help guide the next steps.

\subhead{Semantics from queries.} 
Getting table understanding models to work in practice does not only impose challenges but also presents opportunities. An important aspect of semantics that prior deep representation learning research for both languages and tables ignored is the action or interactive semantics. A table’s semantics is not only embodied by its values and surrounding context (schema, title, etc.) but also by what people (users) do with the table. An important class of actions on tables is SQL queries run by users on tables. We believe that deploying table understanding models within enterprise applications such as Sigma opens up opportunities to incorporate semantics implied by queries into table understanding models. 

\section{Conclusion}\label{sec:conclusion}
At Sigma Computing, we are building the technical foundation to integrate table understanding models with our product that will improve almost every aspect of user experience. Here we introduce a facet of this vision.
Table understanding models have been demonstrated to yield promising performance for detecting column types, entities, relations, and table topics. We believe the time has come to make these advances accessible to the masses, and explore avenues to make table understanding models work in practice. 

To this end, we discuss the main challenges of deploying these models, being: unknown data domains, unrepresentative training data, and unreliable inferences. We propose a general framework to address these challenges in any table understanding system. We present \system which implements this framework. To accomplish model customization, we leverage feedback and use the data programming by demonstration framework to infer labeling functions from data used to generate new training data and prediction functions. \system is trained on data common in enterprise data warehouses and implements a hybrid and adaptive model and detects out-of-distribution data.

The framework introduces many avenues for future research. We are particularly interested in methods to adapt and enrich models based on the user's explicit feedback and implicit signals. Besides the adaptive capabilities for table understanding models, we need to better understand how to best represent tables and develop models accordingly.

\newpage

\bibliographystyle{ACM-Reference-Format}
\bibliography{main}

\end{document}